\documentclass{ptptex}

\usepackage{graphicx}

\newcommand{\la}{\langle}
\newcommand{\ra}{\rangle}

\title{
Long-distance behavior of $q\bar{q}$ color dependent potentials
at finite temperature 
}

\author{
Atsushi \textsc{Nakamura} and
Takuya \textsc{Saito}
}

\inst{
Research Institute for Information Science and Education,
Hiroshima University, Higashi-Hiroshima 739-8521, Japan
}

\abst{
$SU(3)$ heavy quark potentials at finite temperature
are studied with quenched lattice QCD using the stochastic gauge-fixing method.
In addition to the standard color average channel,
we investigate $q\bar{q}$ potentials in singlet and octet channels.
The singlet $q\bar{q}$ channel yields an attractive force, while
the octet $q\bar{q}$ potential is repulsive; the corresponding
color average channel also results in an attractive force.
As the temperature increases, these forces become weak and
their variations are very small; at $T/T_c=1.8-5.6$, 
the singlet attractive force survives over $R \sim 1/T$.
The singlet and octet potentials calculated 
with this algorithm have a small gauge dependence 
when the gauge parameter $\alpha$ is changed from $0.6$ to $1.3$.
 }

\begin{document}
\maketitle
\section{Introduction}

The color screened heavy $q\bar{q}$ potential at finite temperature
plays an important role in the study of $J/\psi$ suppression 
in the quark-gluon plasma (QGP),
\cite{Matsui,Miyamura} 
and has been intensively investigated using lattice simulations
in the two-color \cite{Irback,Heller,Petreczky}
and three-color cases \cite{Attig,Gao,Kaczmarek,Kaczmarek2}.
Quarkonium dissociation by color screening in a deconfining
medium has been studied using the color singlet potential.\cite{Digal}
Fingerprints of QGP have been observed in experiments at CERN-SPS and BNL-RHIC,
 and one expects that
$J/\psi$ suppression and heavy hadron mass modification,
 which are realized in early stages of high energy heavy ion collisions,
 may be a definite evidence of a new form of matter, QGP.
The heavy quark potentials in the QGP phase
can be determined nonperturbatively through
the Polyakov loop correlator (PLC). 
The color average potential has been investigated extensively, 
but the detailed numerical calculation of singlet and adjoint channels,
which give us more useful information on the dynamics,
has just begun to develop;
this is a much more cumbersome task, because we require gauge fixing
for the lattice calculation of the singlet and adjoint
potentials.
Recent progress in lattice QCD techniques and high-speed computers
allow us to study this problem.

The first paper to consider the color dependent $q\bar{q}$ potential
is, to our knowledge, Ref. \citen{McLerran} of McLerran and Svetitsky.
The singlet and adjoint potentials of the SU($N$) potentials, $1$ and
$N^2-1$ (singlet and octet for SU(3)), constructed through PLC, are studied 
in Refs. \citen{Nadkarni} and \citen{Philipsen}.  
In the latter, it is argued that the singlet and adjoint potentials
are gauge independent in a class of the gauge that is local in time.
\footnote{The Lorentz gauge and temporal gauge are not in this class.}
A numerical study in four dimensions was first performed by
Attig et al. \cite{Attig}.
Recently, the Bielefeld group reported
 the results of a singlet potential calculation in
full QCD\cite{Kaczmarek3}; they calculated the ``renormalize Polyakov
loop'' \cite{Kaczmarek2} and extracted the free energy 
by comparing PLC with $T=0$ short-distance behavior.   
In Ref. \citen{MNN03}, the color-dependent PLC was studied at finite
density for the two-color case.
More detailed studies, $i.e.$, which channel is screened and
 the temperature dependence of the screening masses,
 will provide useful information concerning QGP dynamics.
In particular, the singlet channel is important to study 
the heavy quark meson state.

In this letter,
we report lattice QCD study of the temperature dependence of $q\bar{q}$ 
potentials in the deconfinement phase in the quenched approximation. 
Using a stochastic gauge-fixing quantization (SGFQ)
with a Lorentz-type gauge-fixing term,  
\cite{Mizutani} we obtain the color singlet and octet potentials
as well as the color average one.
In the previous studies of color-dependent PLC, 
 a gauge fixing technique based on 
the iteration method is used and the gauge dependence of the singlet and 
octet parts was not investigated. 
In our study, using the SGFQ method with Lorentz-type gauge fixing,
 we find a clear signal of three types of potentials and investigate
the temperature dependence of these potentials and the screening
mass.
We vary the gauge parameter $\alpha$ in SGFQ
 and investigate the gauge parameter $\alpha$ dependence
of the results. 

\section{Heavy quark potentials}

The heavy quark potential can be defined in terms of a Polyakov loop,
\begin{equation}
 L(R) = \prod_{t=1}^{N_t} U_{0}(R,t),
\end{equation}
where the $U_{\mu}$ are lattice link variables.
The expectation value $\la\mbox{Tr}L(0)\ra$
is the order parameter of the confinement/deconfinement
phase transition in quenched QCD \cite{McLerran}.
Using the Polyakov loop, we here construct the PLC
and study three kinds of $q\bar{q}$ potentials.\footnote{Strictly speaking,
the Polyakov line correlations give the excess free energy, $F = U - TS$.}
 \cite{Nadkarni}
The decomposition of the color $SU(3)$ symmetry,
$ 3 \otimes 3^{*} = 1 \oplus 8 $,
yields the singlet and octet potentials.
The color average potential is given by the sum of 
the singlet and octet channels:
\begin{equation}
\begin{array}{cll}
\displaystyle
\exp(-\frac{V_{1}^{q\bar{q}}(R)}{T})
&=&\displaystyle
3\frac{\la\mbox{Tr}L(R)L^{\dagger}(0)\ra}
{\la\mbox{Tr}L(0)\ra^2},\\
\displaystyle
\exp(-\frac{V_{8}^{q\bar{q}}(R)}{T})
&=&\displaystyle
\frac{9}{8}\frac{\la\mbox{Tr}L(R)\mbox{Tr}L^{\dagger}(0)\ra}
{\la\mbox{Tr}L(0)\ra^2}
-\frac{3}{8}\frac{\la\mbox{Tr}L(R)L^{\dagger}(0)\ra}{\la\mbox{Tr}L(0)\ra^2},\\
\displaystyle
\exp(-\frac{V_{c}^{q\bar{q}}(R)}{T})
&=&\displaystyle
\frac{\la\mbox{Tr}L(R)\mbox{Tr}L^{\dagger}(0)\ra}
{\la\mbox{Tr}L(0)\ra^2}\\
&=& \displaystyle
\frac{1}{9}\exp(-\frac{V_1^{q\bar{q}}(R)}{T})
 + \frac{8}{9}\exp(-\frac{V_8^{q\bar{q}}(R)}{T})\label{qqbp}.
\end{array}
\end{equation}
All potentials in Eq. (\ref{qqbp}) are functions of the temperature, $T$.
In this study, because we focus on a finite temperature simulation
 with $\la\mbox{Tr}L(0)\ra\neq 0$,
the above PLCs are normalized
by the value of $\la\mbox{Tr}L(0)\ra^2$.

In the leading-order perturbation (LOP), $V_1^{q\bar{q}}$ and 
$V_8^{q\bar{q}}$ can be classified
according to the coefficient of the color exchange dynamics:
$C_{q\bar{q}}[1]=-\frac{4}{3}$ and $C_{q\bar{q}}[8]=+\frac{1}{6}$,
 respectively.
The singlet potential $V_1^{q\bar{q}}$ yields an attractive force,
which has a close relation to the mesonic bound states, while
 $V_8^{q\bar{q}}$ yields a repulsive force,
 which weakens the binding force among quarks
at finite temperature.
The resultant $V_c^{q\bar{q}}$ is generally attractive.

\section{Stochastic gauge fixing}

The potentials except for $V_c^{q\bar{q}}$
are not gauge invariant, and accordingly a gauge fixing procedure is needed;
however, the lattice gauge fixing by an iteration method,
 most commonly used in the literature \cite{Mandula},
becomes cumbersome as the lattice becomes large.
In fact, this is the main reason why the lattice study of 
 gauge non-invariant heavy
quark potentials has not been carried out more extensively.
We here employ the SGFQ with the Lorentz-type gauge fixing
proposed by Zwanziger\cite{Zwanziger}:
\begin{equation}
\frac{dA_{\mu}^a}{d\tau}=
-\frac{\delta S}{\delta A_{\mu}^a } +
\frac{1}{\alpha} D_{\mu}^{ab}(A)
\partial_{\nu} A_{\nu}^b + \eta_{\mu}^a\label{sq},
\end{equation}
where
$\alpha$, $D_{\mu}^{ab}$ and $\eta_{\mu}^a$ represent a gauge parameter,
a covariant derivative and Gaussian random noise, respectively.
This algorithm enables us to perform a lattice gauge fixing simulation
without iteration and may be suitable for the calculation of
$V_1^{q\bar{q}}$ and $V_8^{q\bar{q}}$, 
which may require a relatively large amount of data.
Because the parameter $\alpha$ can be changed for each simulation
the method is useful for examining the gauge $\alpha$-parameter dependence
of $V_1^{q\bar{q}}$ and $V_8^{q\bar{q}}$.
The lattice formulation and a more detailed explanation
 of this algorithm can be found
 in Refs. \citen{Mizutani}, \citen{Saito} and \citen{Saito2}.

\section{Numerical result}

We carried out a quenched $SU(3)$ lattice simulation
using the standard plaquette action.
The spatial volume was $24^3$ and
the temporal lattice size was set as $N_t= 6$,
which determines the temperature as $T=1/N_ta$, 
where $a$ is the lattice cutoff. 
The critical temperature is 
estimated as $T_c \sim 256 \mbox{ MeV}$ ($N_t=6$) in Ref. \citen{Boyd}.
The system temperature changes when we vary the lattice cutoff;
the relationship between the lattice coupling constant and the cutoff
is determined by Monte Carlo renormalization group analysis \cite{QCD_TARO}.
All of the PLCs were measured every ten Langevin steps.
The amount of statistics is from 3000 to 10000
after approximately 3000 steps, regarded as the thermalization process, 
 were discarded.
\footnote{As the system temperature decreases, 
 more data are required.}
 The value of $\alpha$ was fixed to $1.0$, 
 except when we studied the gauge dependence of the potentials.

\begin{figure}[htb]
\begin{center}
\resizebox{14cm}{!}
{\includegraphics{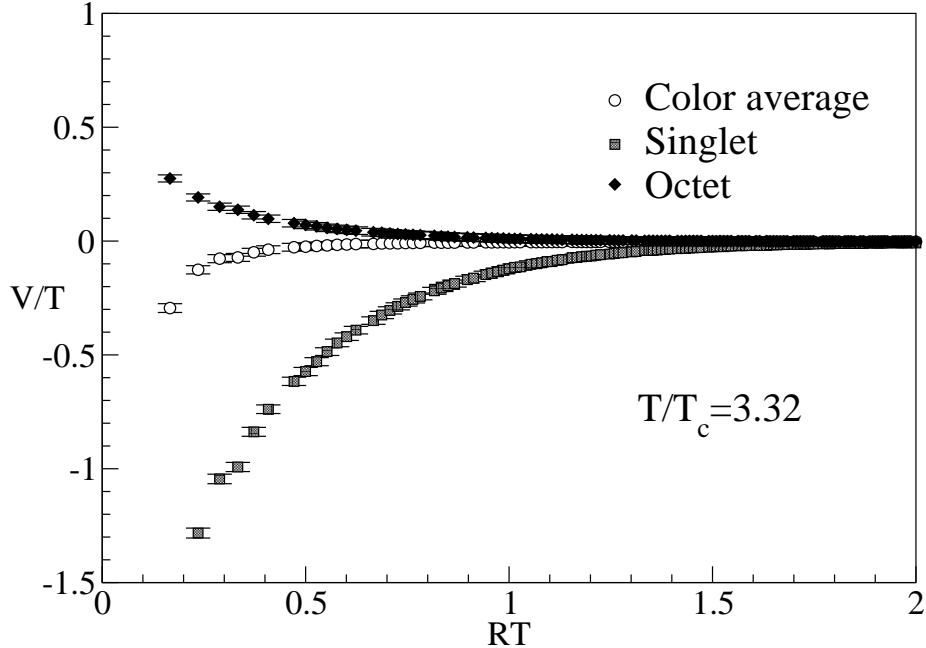}}
\caption{The potential as a function of $RT$ in different channels
at $T/T_c=3.32$. Here, $\Delta \tau = 0.03$.}\label{qqb-zt}
\end{center}
\end{figure}

In Fig. \ref{qqb-zt},
we show typical behavior of three kinds of potentials
at $T/T_c=3.32$.
As expected from consideration of the color exchange dynamics,
$V_1^{q\bar{q}}$ yields an attractive force, while
$V_8^{q\bar{q}}$ yields a repulsive force.
\footnote{The same behavior for the $SU(3)$ $q\bar{q}$
potentials is reported in Ref. \citen{Attig}.}
This repulsive potential is weaker than
the attractive potential,
and correspondingly $V_c^{q\bar{q}}$ also yields an attractive force.
At large $R$, each potential becomes flat, and
 the effective force due to the color screening 
becomes zero.
$V_c^{q\bar{q}}$ reaches the flat region most rapidly, 
and then next does the $V_8^{q\bar{q}}$ repulsive force.
Both of these vanish at $RT<1$.
$V_1^{q\bar{q}}$ becomes flat at larger distances, beyond $RT=1$.
We note that the singlet channel 
controls heavy meson spectroscopy in the QGP.

\begin{figure}[htb]
\begin{center}
\resizebox{14cm}{!}
{
\includegraphics{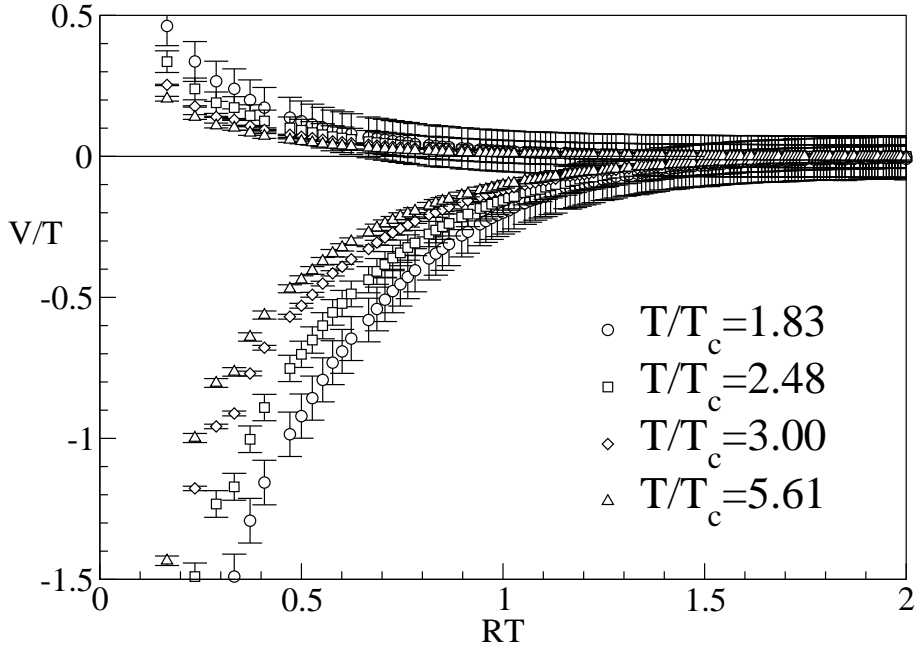}
}
\caption{Temperature dependence of the singlet and octet potentials. 
Here $\Delta \tau = 0.03$.}\label{t-deps}
\end{center}
\end{figure}

In order to investigate the $T$ dependence of the potentials,
we present $V/T$ in the color singlet and octet channels 
at $T/T_c=1.83, 2.48, 3.00$ and $5.61$ in Fig. \ref{t-deps}.
As $T$ increases, the flat region
in each potential moves to shorter distances.
Even at very high temperatures, for example, $T/T_c=5.61$, 
the flat region of the singlet potential is above 
$R=1/T$.
The singlet channel possesses a long-distance attractive force
after the deconfinement phase transition.
%

\begin{figure}[htb]
\begin{center}
\resizebox{14.0cm}{!}
{\includegraphics{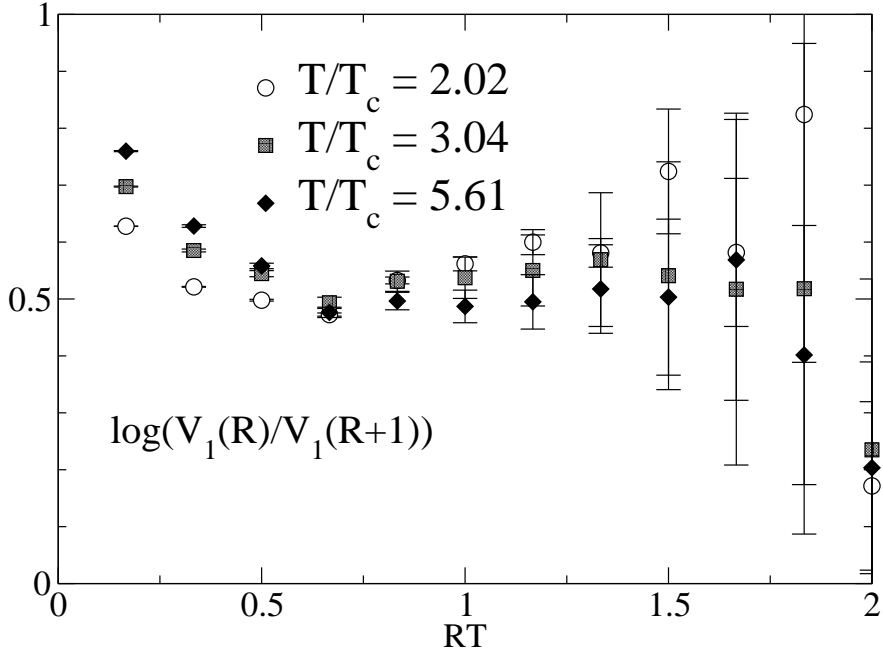}}
\caption{Effective screening mass for the singlet potentials.
 Here $\Delta \tau = 0.03$.}
\label{effm}
\end{center}
\end{figure}

\begin{figure}[htb]
\begin{center}
\resizebox{14.0cm}{!}
{\includegraphics{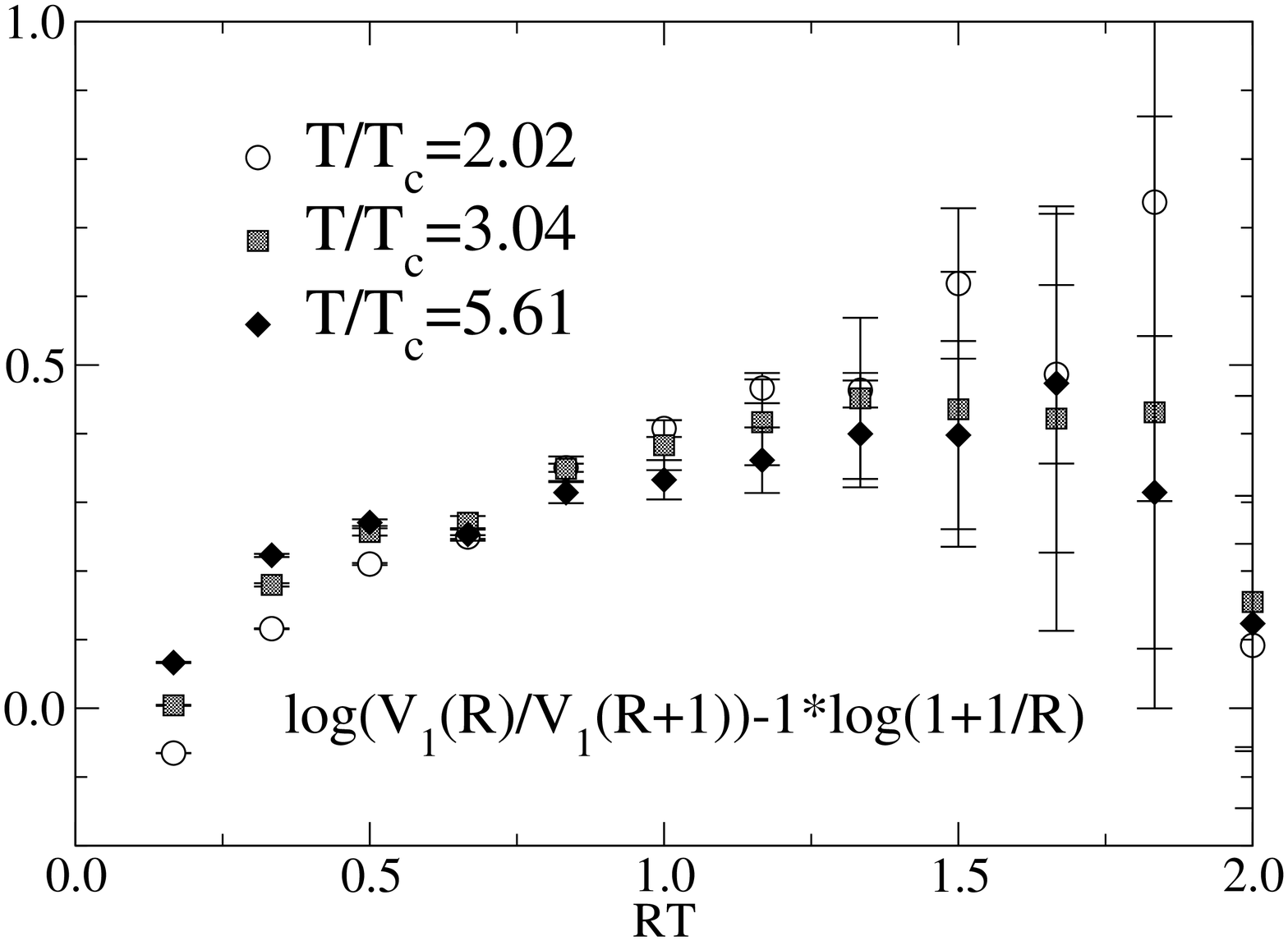}}
\caption{Effective screening mass for the singlet potentials.
The screening masses are defined as 
$\log(V_1(V(R)/V_1(R+1))-1\times \log(1+1/R)$. Here $\Delta \tau = 0.03$.}
\label{effm2}
\end{center}
\end{figure}

The extraction of the screening (electric) mass,
which is an essential factor in QGP physics, is very important. 
The screened $q\bar{q}$ potential 
is expected to have the form  
\begin{equation}
\frac{V(R,T)}{T} = C(T) \frac{\exp(-m_e(T)R)}{R^{d}},
\end{equation}
where $m_e(T)$ is the screening mass.
Then, we have 
\begin{equation}
\log(\frac{V(R)}{V(R+1)}) = m_e + d \log\left(1+\frac{1}{R}\right).
\label{m-eff}
\end{equation}
If the distance is sufficiently large, then Eq. (\ref{m-eff})
reaches the screening mass, $m_e$. 
Numerical data of $\log(V(R)/V(R+1)$ for the singlet channel  
are shown in Fig. \ref{effm}, and the values of $m_e$ in the case $d=1$
are plotted on Fig. \ref{effm2}.
At short distances, they decrease, and become flat 
at large distances, $RT>1$.
The long-distance value of the mass at $T/T_c=2.02$ is slightly larger than
that at the higher temperature, $T/T_c=5.61$, and the variation 
is not large in these temperature regions.
This result is consistent with the previous screening mass study 
in which the $SU(3)$ gluon screened propagators
 are computed. \cite{Saito,Saito2}

\begin{figure}[htb]
\begin{center}
\resizebox{14.0cm}{!}
{\includegraphics{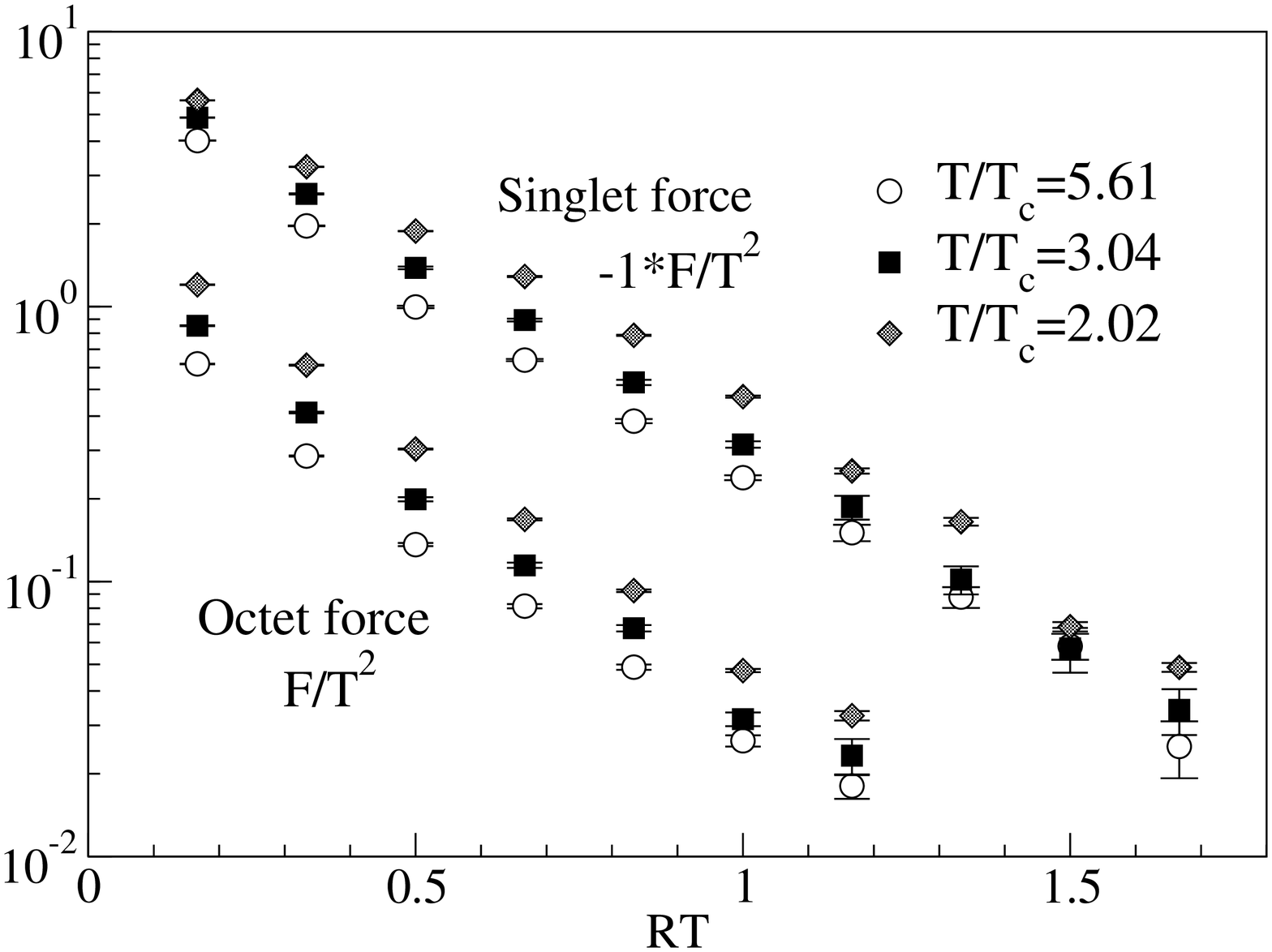}}
\caption{Effective forces from the singlet and octet channels.
 Here, $\Delta \tau = 0.03$.}
\label{eforce}
\end{center}
\end{figure}

We can define an effective force from the singlet and octet potentials as
\begin{equation}
\frac{F}{T^2} = -\frac{1}{T}(V(R+1)-V(R)).
\end{equation}
The temperature dependence of this force is shown in Fig. \ref{eforce}.
The octet force is smaller than the singlet force
at all distances, and at long distances, the value of
the effective force approaches zero. 
As the temperature increases, both singlet and octet forces
become weaker at all distances.

\begin{figure}[htb]
\begin{center}
\resizebox{14.0cm}{!}
 { \includegraphics{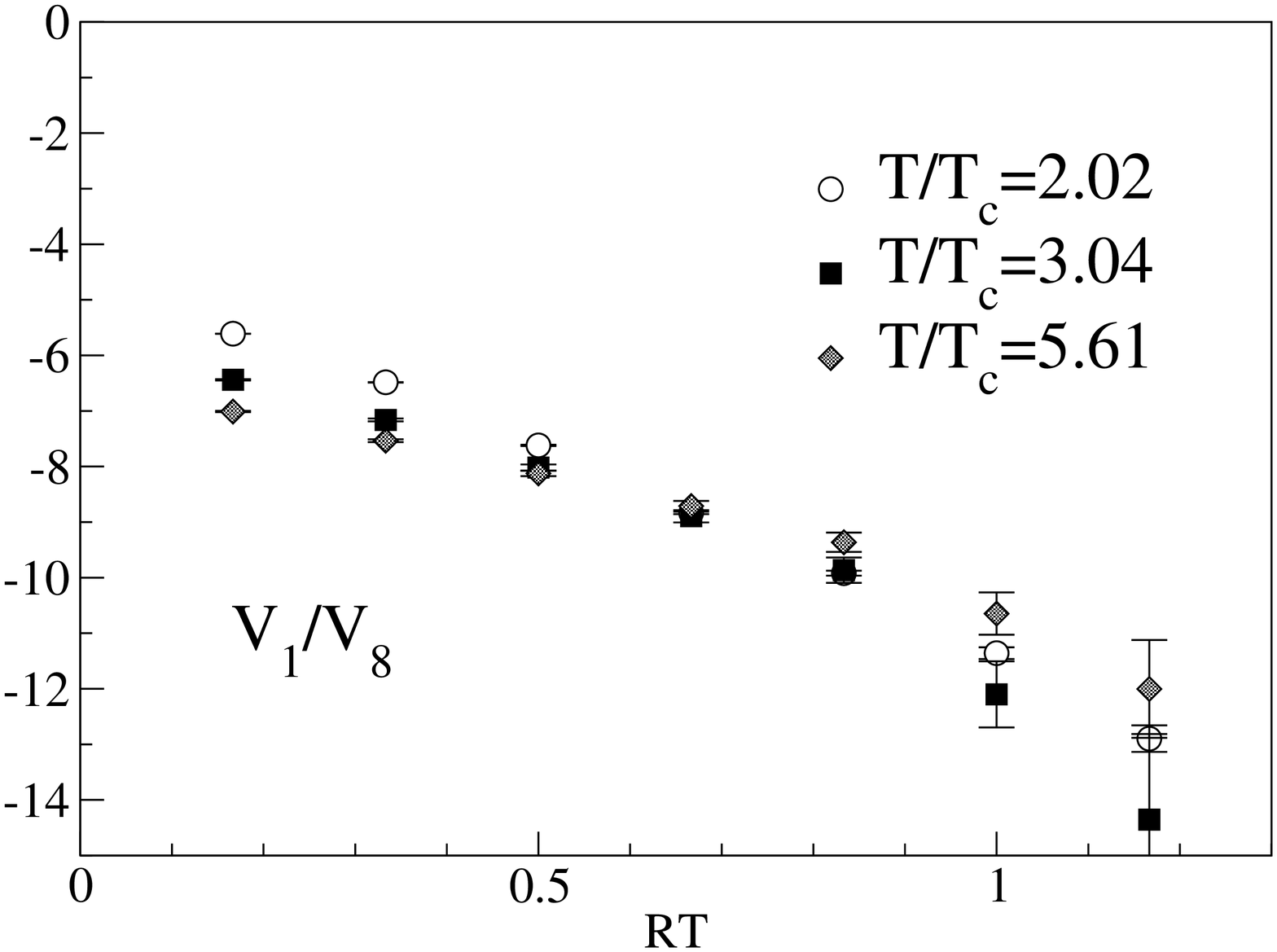} }
\caption{Ratio $V_1/V_8$, at short distance region.
Here, $\Delta \tau = 0.03$.
\vspace{0.5cm}
}\label{ratio}
\end{center}
\end{figure}

In order to compare the strength of the two forces, 
we define the ratio $V_1^{q\bar{q}}/V_8^{q\bar{q}}$;
 it approaches the value $C_{q\bar{q}}[1]/C_{q\bar{q}}[8] = -8$, at
short distances, as predicted by LOP
\cite{McLerran,Nadkarni}.
Figure \ref{ratio} displays the computed ratios
at $T/T_c=2.02, 3.04 \mbox{ and } 5.61$.
We find that as $T$ increases, the ratio is comparable with
 the expected values;
however, as $T$ decreases, it deviates from $-8$.
There are several possible causes for this deviation at shorter distances
 in low $T$.
One is numerical instability due to the long autocorrelation near $T_c$
and thus larger scale simulations in the vicinity of $T_c$ are required.
Another is the renormalization of PLC;
i.e., $\langle Tr L(R) L^\dagger(0) \rangle = Z \exp(-F(R)/T)$,
 and if $Z$ differs greatly from 1, 
 we should take it into account when comparing $F$
with that obtained by the perturbation.
Recently, the Bielefeld group proposed that the proper normalization of PLC
can be determined by the comparison of the $T=0$ potential data with
the corresponding $T\neq0$ singlet free energy at short distances
\cite{Kaczmarek2}.
However, the determination of the normalization of the octet channel
using this method was not made.
The third possibility is that our simulations on the $N_t=6$ lattice
 are not sufficiently close to the continuum 
 and that lattice artifact makes the comparison with the
continuum perturbation difficult even at short distances.

\begin{figure}[htb]
\begin{center}
\resizebox{14.0cm}{!}
{ \includegraphics{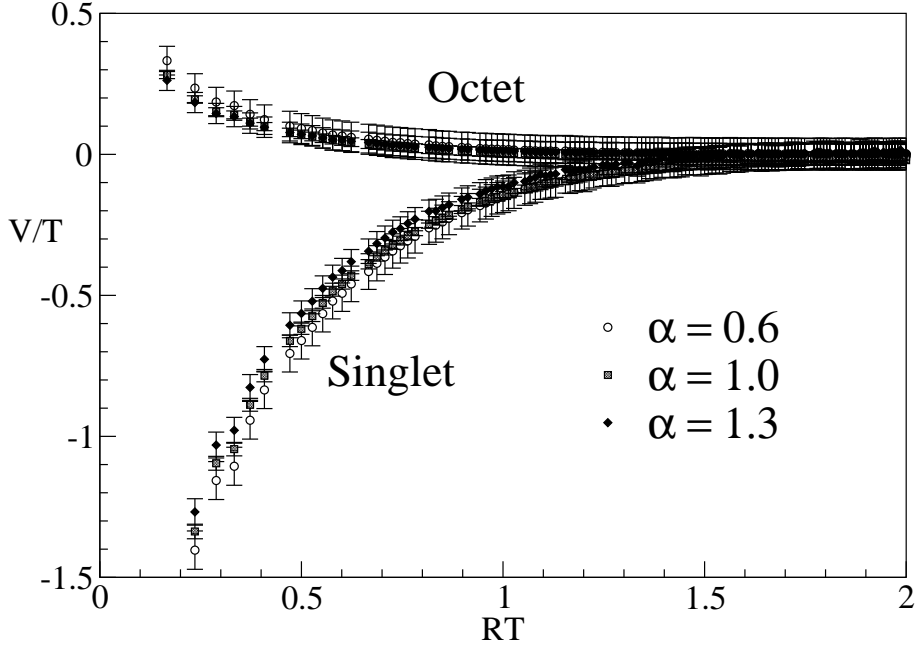} }
\caption{Gauge parameter $\alpha$ dependence
 at $T/T_c=3.04$ ($\Delta \tau =0.03$ ).}
\label{dog}
\end{center}
\end{figure}

\begin{figure}[hbt]
\begin{center}
\resizebox{14.0cm}{!}
{\includegraphics{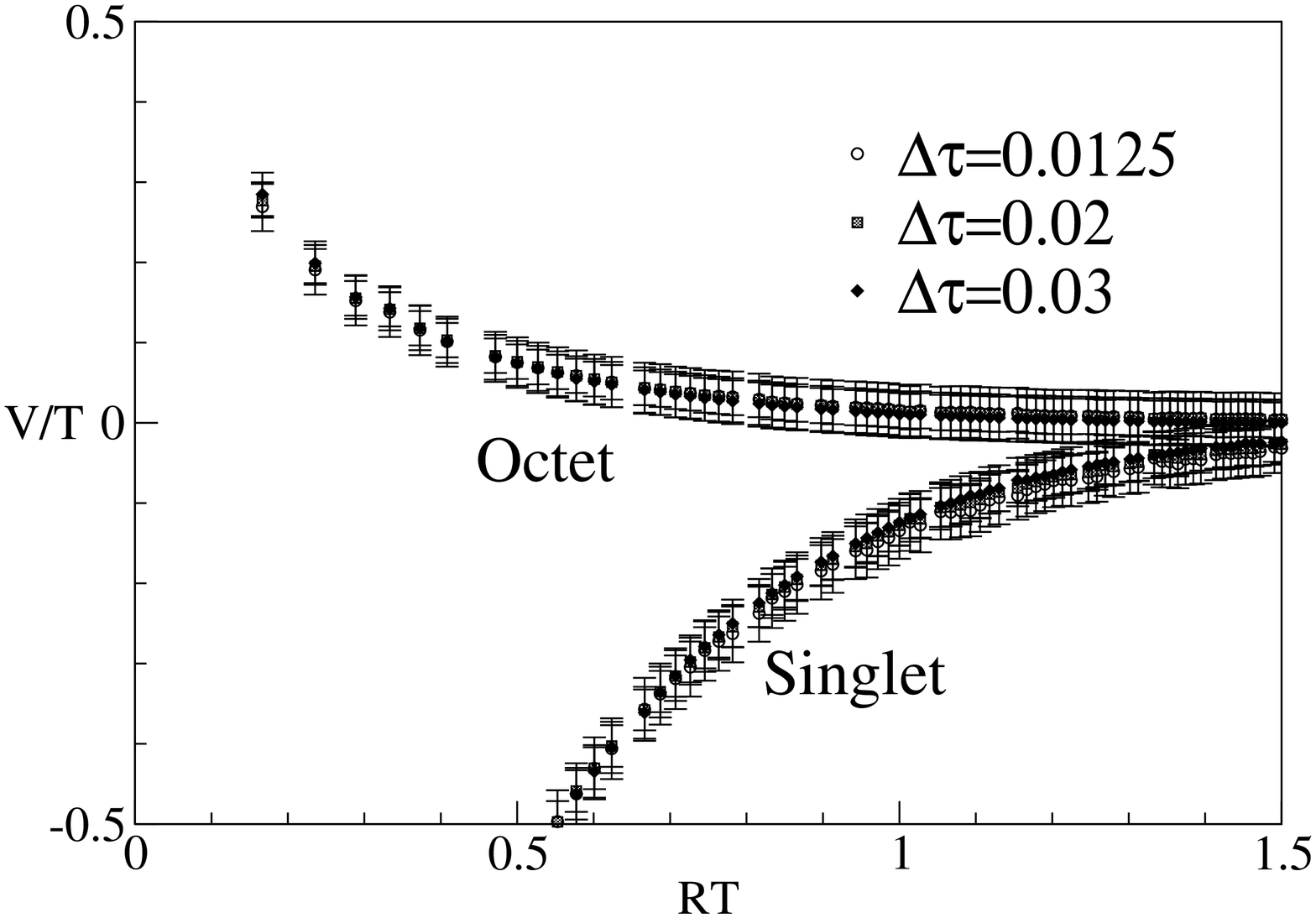}}
\end{center}
\caption{Langevin step width $\Delta \tau$ dependence of the singlet
 and octet potentials at $T/T_c=3.04$.}\label{dodt}
\end{figure}

Although $V_1^{q\bar{q}}$ and $V_8^{q\bar{q}}$
are determined in the framework of the SGFQ,
they are not gauge invariant.
It is therefore very important to investigate their gauge $\alpha$-parameter 
dependence.
Figure \ref{dog} displays the gauge parameter dependence
of $V_1^{q\bar{q}}$ as $\alpha$ is varied from 0.6 to 1.3.
We see that this dependence is small, and their basic features do not change.

We solve Eq. (\ref{sq}) with finite $\Delta \tau$, and
to obtain final reliable data,
we must study the $\Delta \tau$ dependence.
In Fig. \ref{dodt}, where $V_1^{q\bar{q}}$ and $V_8^{q\bar{q}}$
are plotted for several values of $\Delta \tau$,
their behavior is not strongly affected
for $\Delta \tau =0.0125-0.03$.
Note that we use the Runge-Kutta algorithm to reduce
the $\Delta \tau$ dependence \cite{Batrouni}.

\begin{figure}[htb]
\begin{center}
\resizebox{14.0cm}{!}
{ \includegraphics{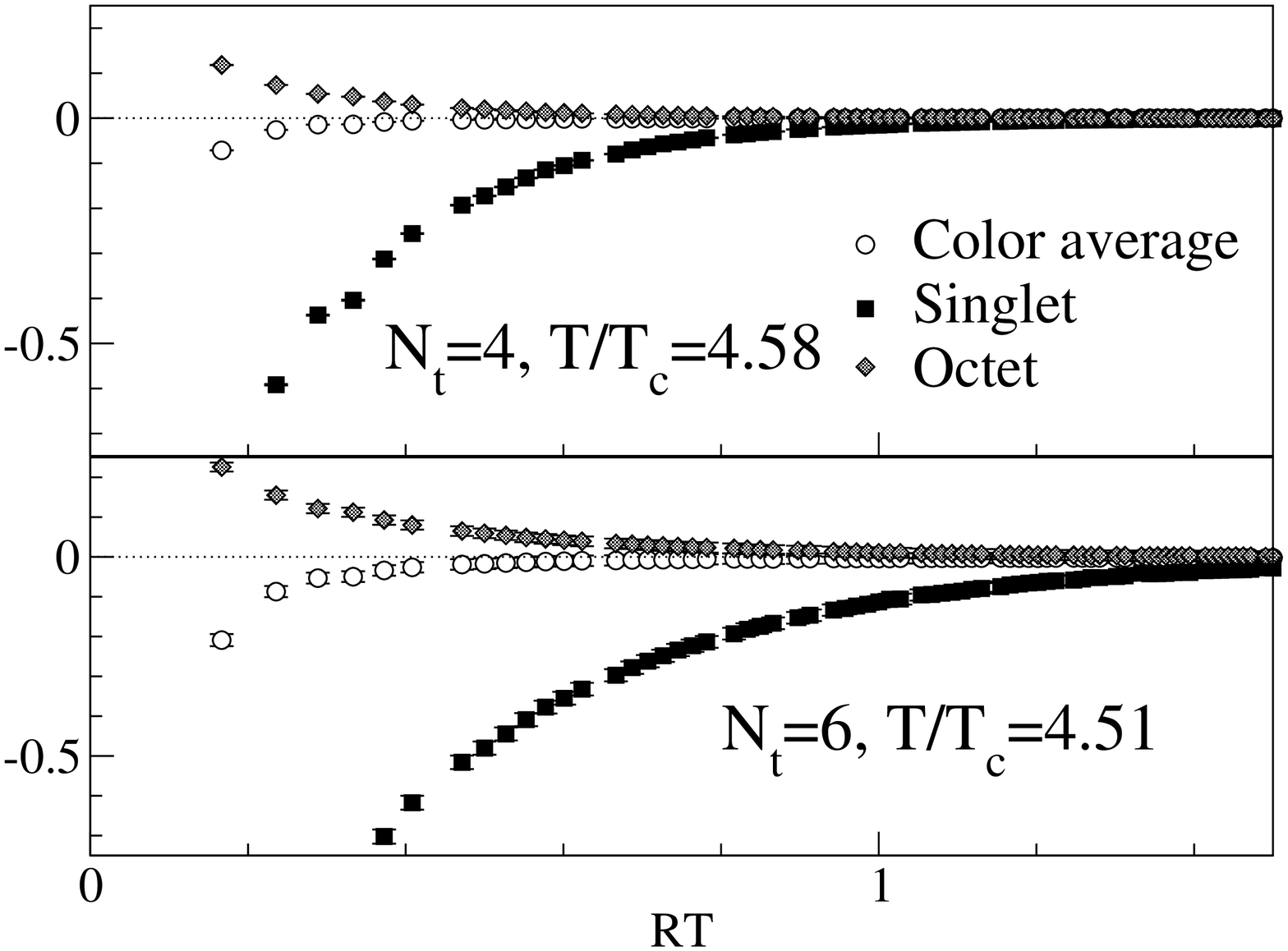} }
\caption{Simulation results for $N_t=4$ and $N_t=6$ lattices.
Here $\Delta \tau = 0.03$.
}\label{comp}
\end{center}
\end{figure}

These lattice calculations were carried out on a $N_t=6$ lattice.
If we calculate the PLC using a $N_t=4$ lattice simulation, 
 then we obtain different results in comparison with the $N_t=6$ case.
In Fig. \ref{comp} we find that the flat region of the singlet potential
from the $N_t=4$ lattice simulation is located at shorter distances than
that in the $N_t=6$ case. 
The lattice simulation of the equation of state reported in Ref. \citen{Boyd}
suggests that the data from the $N_t=4$ lattice calculations are 
far from the continuum limit.
The studies reported in Ref. \citen{Attig} were performed
 on $12^3 \times 4$ lattices, and
our simulation results here provide the $SU(3)$ singlet and octet 
potential nearer the continuum.

\section{Concluding remarks}

We have studied the $q\bar{q}$ potentials
at finite temperature above $T_c$ through lattice quenched $SU(3)$ simulations.
In order to calculate $V_1^{q\bar{q}}$ and $V_8^{q\bar{q}}$, 
 we employed SGFQ and clearly obtained 
the expected behavior for three kinds of potentials in the QGP phase;
namely, $V_1^{q\bar{q}}$ yields an attractive force,
$V_8^{q\bar{q}}$ a repulsive force, and $V_c^{q\bar{q}}$ an attractive force.
The increase of $T$ weakens the forces produced by these potentials, 
and the singlet potential yields a long-distance attractive force
 even at $T/T_c=1.8-5.6$, which corresponds to
 RHIC and LHC heavy ion collision experiments.
These results suggest that QGP has a nontrivial structure and
is not a free gas.
Recent lattice studies of the charmonium bound state
at finite temperature employing the maximum entropy method \cite{MEM} 
show the possibility that $J/\psi$ survives at high temperatures.
Moreover, a simple perturbative prediction for gluon electric and magnetic
screening masses cannot be applied in the QGP system
\cite{Saito,Saito2}. We need more detailed studies
of the heavy quark potential by lattice QCD simulations.

In the SGFQ with a Lorentz-type gauge fixing term, we confirm that
the $\alpha$-parameter dependence of $V_1^{q\bar{q}}$ and $V_8^{q\bar{q}}$ 
is small
and that the qualitative behavior of these potentials is not changed 
when $\alpha$ is varied from $0.6$ to $1.3$.
This is the first step in the gauge dependence study of the
potentials calculated by the color dependent PLC.
We plan more extensive simulations with other gauge fixing methods.
In particular, we will study Coulomb gauge, 
since certain advantages of Coulomb-type gauge fixing have
recently been reported \cite{Philipsen,Zwanziger2}.

We are also interested in the more precise determination 
of the Debye screening mass and its temperature dependence. 
The heavy quark potential in the QGP medium
 is closely related to the gluon screening effect, and
it is important to extract the electric screening mass
from the Polyakov loop correlation functions.

\section*{Acknowledgements}
We would like to thank T. Inagaki for many helpful discussions.
Our calculations
were carried out on a SX-5 (NEC) vector-parallel computer.
We appreciate the warm hospitality and
support of the RCNP and CMC administrators.
This work is supported by Grants-in-Aid for Scientific Research from
the Ministry of Education, Culture, Sports, Science and Technology,
Japan (No. 11694085, No. 11740159, and No. 12554008).

%


\begin{thebibliography}{99}
  

\bibitem{Matsui} T. Matsui and H. Satz,
Phys. Lett. B {\bf 178}, 416 (1986).

\bibitem{Miyamura} T. Hashimoto, K. Hirose, T. Kanki and O. Miyamura,  
	Phys. Rev.Lett. 57,2123 (1986). 

\bibitem{Irback} A. Irb$\ddot{\mbox{a}}$ck,
P. Lacock, D. Miller, B. Petersson and T. Reisz,
Nucl. Phys. B {\bf 363}, 34 (1991).

\bibitem{Heller} U.M. Heller, F. Karsch and J. Rank,
Phys. Rev. D {\bf 57}, 1438 (1998).

\bibitem{Petreczky} S. Digal, S. Fortunato and P. Petreczky,
hep-lat/0304017.

\bibitem{Attig} N. Attig, F. Karsch, B. Petersson, H. Satz and M. Wolff,
Phys. Lett. B {\bf 209}, 65 (1988).

\bibitem{Gao} M. Gao, Phys. Rev. D {\bf 41}, 626 (1990).

\bibitem{Kaczmarek} O. Kaczmarek, F. Karsch, E. Laermann
and M. L$\ddot{\mbox{u}}$tgemeier, Phys. Rev. D {\bf 62}, 034021 (2000).

\bibitem{Kaczmarek2} O. Kaczmarek, F. Karsch, P. Petreczky and F. Zantow,
Phys. Lett. B {\bf 543}, 41 (2002).

\bibitem{Digal} S. Digal, P. Petreczky and H. Satz,
	Phys. Rev. D {\bf 64}, 094015 (2001).

\bibitem{McLerran} L. D. McLerran and B. Svetitsky, Phys. Rev. D {\bf 24},
450 (1981).

\bibitem{Nadkarni} S. Nadkarni, 
	Phys. Rev. D {\bf 34}, 3904 (1986).

\bibitem{Philipsen} O. Philipsen, Phys. Lett. B {\bf 535}, 138 (2002).

\bibitem{Kaczmarek3} O. Kaczmarek, S. Ejiri, F. Karsch, E. Laermann and
F. Zantow, hep-lat/0312015 to appear in the Proceedings of Finite Density
QCD 03, Prog. Theor. Suppl. 

\bibitem{MNN03} S. Muroya, A. Nakamura and C. Nonaka, 
	hep-lat/0208006 to appear in Nucl. Phys. B (Proc. Suppl.) 119 (2003). 

\bibitem{Mizutani} A. Nakamura and M. Mizutani,
{\it Vistas in Astronomy} (Pergamon Press), vol. {\bf 37}, 305 (1993);
M. Mizutani and A. Nakamura, Nucl. Phys. B(Proc. Suppl.) {\bf 34}, 253 (1994);
A. Nakamura, Prog. Theor. Phys. Suppl. No. {\bf 131}, 585, 1998.

\bibitem{Mandula} K.G. Wilson, {\it in Recent Developments in Gauge Theories,}
ed. G. t'Hooft (Plenum Press, New York, 1980), 363;
J.E. Mandula and M. Ogilvie B {\bf 185}, 127 (1987).

\bibitem{Zwanziger} D. Zwanziger, Nucl. Phys. {\bf B192}, 259 (1981).

\bibitem{Saito} A. Nakamura, I. Pushkina, T. Saito and S. Sakai, 
	Phys. Lett. B {\bf 549}, 133 (2002).
\bibitem{Saito2} A. Nakamura, T. Saito and S. Sakai,
        Phys. Rev. D {\bf 69}, 014506 (2004).

\bibitem{Boyd} G. Boyd, J. Engels, F. Karsch, E. Laermann, C. Legeland,
M. L$\ddot{\mbox{u}}$tgemeier, and B. Petersson,
Phys. Rev. Lett. {\bf 75}, 4169 (1995); Nucl. Phys. B {\bf 469}, 419 (1996).

\bibitem{QCD_TARO} QCDTARO collaboration, K. Akemi, et al.,
Phys. Rev. Lett. {\bf 71}, 3063 (1993).




\bibitem{Batrouni} A. Ukawa and M. Fukugita, Phys. Rev. Lett. {\bf 55}, 1854
(1985); G.G. Batrouni, et al., Phys. Rev. D {\bf 32}, 2736 (1985).


\bibitem{MEM} M. Asakawa and T. Hatsuda, Phys. Rev. Lett. 92 (2004) 012001; 
T. Umeda, K. Nomura, and H. Matsufuru, hep-lat/0211003;
S. Datta, F. Karsch, P. Petreczky, I. Wetzorke, hep-lat/0312037.


\bibitem{Zwanziger2} D. Zwanziger, Phys. Rev. Lett. {\bf 90}, 102001 (2003).

\end{thebibliography}
\end{document}